\documentclass[sigconf, screen, authorversion]{acmart}

\copyrightyear{2022}
\acmYear{2022}
\setcopyright{acmlicensed}\acmConference[CUI 2022]{4th Conference on Conversational User Interfaces}{July 26--28, 2022}{Glasgow, United Kingdom}
\acmBooktitle{4th Conference on Conversational User Interfaces (CUI 2022), July 26--28, 2022, Glasgow, United Kingdom}
\acmPrice{15.00}
\acmDOI{10.1145/3543829.3544513}
\acmISBN{978-1-4503-9739-1/22/07}

\begin{document}

\title[A Provocation on the Certification of Skills in the Alexa and Google Assistant Ecosystems]{When It's Not Worth the Paper It's Written On: A Provocation on the Certification of Skills in the Alexa and Google Assistant Ecosystems}

\author{William Seymour}
\email{william.1.seymour@kcl.ac.uk}
\affiliation{%
  \institution{King's College London}
  \streetaddress{Bush House, 30 Aldwych}
  \city{London}
  \country{UK}
  \postcode{WC2B 4BG}
}
\author{Mark Cot\'{e}}
\email{mark.cote@kcl.ac.uk}
\affiliation{%
  \institution{King's College London}
  \streetaddress{Chesham Building, Strand}
  \city{London}
  \country{UK}
  \postcode{WC2R 2LS}
}
\author{Jose Such}
\email{jose.such@kcl.ac.uk}
\affiliation{%
  \institution{King's College London}
  \streetaddress{Bush House, 30 Aldwych}
  \city{London}
  \country{UK}
  \postcode{WC2B 4BG}
}

\begin{abstract}
The increasing reach and functionality of voice assistants has allowed them to become a general-purpose platform for tasks like playing music, accessing information, and controlling smart home devices. In order to maintain the quality of third-party skills and to protect children and other members of the public from inappropriate or malicious skills, platform providers have developed content policies and certification procedures that skills must undergo prior to public release. Unfortunately, research suggests that these measures have been ineffective at curating voice assistant platforms, with documented instances of skills with significant security and privacy problems. This provocation paper outlines how the underlying architectures of these platforms had turned skill certification into a seemingly intractable problem, as well as how current certification methods fall short of their full potential. We present a roadmap for improving the state of skill certification on contemporary voice assistant platforms, including research directions and actions that need to be taken by platform vendors. Promoting this change in domestic voice assistants is especially important, as developers of commercial and industrial assistants or other similar contexts increasingly look to these devices for norms and conventions. 
\end{abstract}



\maketitle

\section{Introduction}
Voice assistant (VA) platforms have evolved considerably since their inception, growing from a handful of available skills to hundreds of thousands~\cite{edu2020smart}. As the use of voice assistants has similarly increased, users have become more concerned about the privacy, safety, and security of the both voice assistants as products and the skills that run on them~\cite{abdi2019more,abdi2021privacy}. Focusing on the latter, while major platforms such as Amazon and Google have policies intended to protect users, research has shown that contravening skills are nonetheless present on skill stores for people to use~\cite{edu2021skillvet, le2021skillbot, guo2020skillexplorer, young2022skill,edu2022measuring}.

While platforms have certification procedures in place to ensure that skills comply with policies, it is becoming increasingly clear that current practices are not sufficient to protect users by keeping disallowed skills from marketplaces. The introduction of legislation aimed at protecting children and personal data online\footnote{Such as the Children's Online Privacy Protection Act and the EU General Data Protection Regulation}, has taken place during a gradual shift in political sentiment that increasingly requires platform operators to vet content submitted by third parties. The presence of these devices in the home further introduces concerns around access to speech by developers, with widespread fears and rumours about devices listening into private conversations~\cite{10.1145/3357236.3395501, 10.1145/3274371}.

Unfortunately, the open ended nature of speech and the wide variety of devices that voice assistants can be invoked from makes detecting foul play difficult and architectural constraints on how skills are developed and deployed further increases certification challenges. Work on VA ecosystems highlights how skills violating platform policies (including policies for childrens' skills) make it through current certification processes~\cite{10.1145/3372297.3423339}, and that only a small subset of conversation pathways are actually tested in submitted skills~\cite{10.1145/3478101}. At the same time, skill `stores' (e.g. for Alexa and Google Assistant) give users misplaced confidence in the skills that they use by emulating smartphone app stores and e-commerce platforms. Further comparison with smartphone app stores reveals interesting parallels; early versions of major smartphone app stores were unvetted, with Google arguing that ``open systems win''. Despite this, since 2012 the Android app store has featured an increasing number of verification mechanisms enforcing security, privacy, and content policies for submitted apps. This transition seems to be aligned with the policies of other vendors such as Apple who operate similar closed platforms with sophisticated automated mechanisms to protect users.

This provocation paper outlines how the architecture of the Alexa and Google assistant voice assistant platforms has lead to an intractable certification process by collecting and relating key issues with skill certification and platform architectures. Subsequent discussion shows how these issues come together to make voice assistant platforms so difficult to manage for vendors and developers and highlights improvement opportunities. From this starting point we call on the research community to continue the research agenda by investigating these issues of transparency and accountability, as well as further exploring potential solutions. While the examples in the paper skew towards the Alexa platform due to it being the subject of more research, many of the issues under discussion are symptoms of the underlying architecture of voice assistants and skills used by both; studies that compare the two platforms, such as for certifying skills that violate content policies~\cite{10.1145/3372297.3423339}, show that Amazon \textit{and} Google have much work to do.

\section{Platform Architectures make Certification Infeasible}
\begin{figure}
    \centering
    \includegraphics[width=1.0\columnwidth]{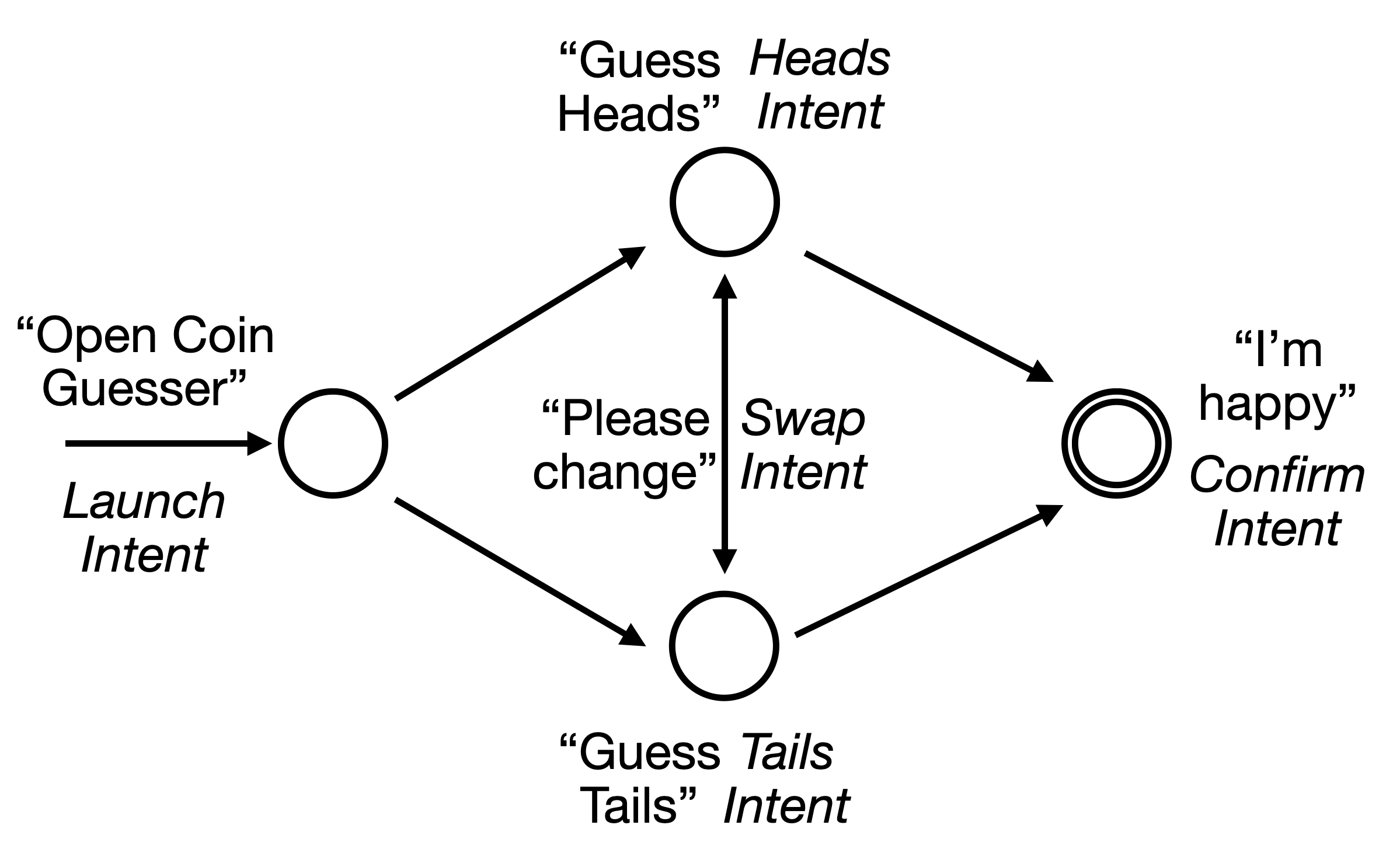}
    \caption{A simple skill represented as a finite state machine with four distinct conversation states and five permitted transitions between states. After the user invokes the skill, they guess the outcome of a coin flip. They can then change their guess as many times as they wish, before confirming and ending the interaction.}
    \label{fig:dfa}
\end{figure}

Place yourself in the perspective of somebody responsible for certifying voice assistant skills and it quickly becomes apparent that they face a challenging task. Because skills can be hosted by developers (rather than on the platform provider's own infrastructure), source code for skills is not included as part of the process to certification and deployment. This means that the only way to check a skill is to interact with it. Platforms do have information about the utterances, intents, and slots that a skill accepts, as these are required to train the recognition models used by the cloud-hosted parts of the VA to route requests, but these lack structure or relation to one another. This is important when interacting with a skill because some intents might only normally be invoked after others (e.g. ``OrderTaxi'' intent after ``SupplyAddress'' intent), resulting in a conversation that behaves like a finite state machine where each state has a set of permitted transitions (see Figure~\ref{fig:dfa}).

Thus, as a human agent there is no option when certifying a skill but to manually invoke it and see which intents it steers the response towards (e.g. ``say yes to start the game, or no to exit'' $\rightarrow$ `Yes' and `No' intents). Simply iterating over the list of registered intents is therefore likely to be inefficient, as many will not function correctly without the correct conversation state, and confusing, as the human agent must jump between different states in the conversation model without the context that may be required to quickly judge whether the skill adheres to platform policies (e.g. whether a referenced product is part of an imaginary game universe or a prohibited advertisement for a real-world item). The skill might also ask the agent a question that is difficult to answer, perhaps because it requires specialist knowledge as in a quiz game.

An automated test script might more efficiently invoke every registered intent in turn and proceed build a conversation tree, mapping out possible conversations from the power set of intents and stopping when the skill ends the interaction. But due to the inability to access source code, such a script would need to interpret the natural language responses from the skill in real-time (indeed, this is what the Amazon and Google automated tests currently do~\cite{10.1145/3478101}). This task is made easier when skills enumerate potential responses, but there are many different ways to ask for a name or address and more open questions may still prove difficult to answer. Supplying many different types of sample data to these intents may help but at the cost of time complexity, and even so, many conversation paths could remain untested if automated mechanisms are unable to invoke the expected intent. Furthermore, skills may be purposely designed to evade this kind of automated testing, similarly to how malware (including on mobile platforms) can increasingly detect when it is being run in a sandbox environment and deactivate to avoid detection~\cite{afianian2019malware}.

Another concern that has been raised in the literature is around access to user's personal data. Normally, developers are required to ask users for permission to access details such as their phone number and location, retrieving these through the platform's own API. This removes the need for developers to store personal data and ensures that data is fresh, as it should be retrieved every time it is used. However, rather than using platform-provided mechanisms to access personal data, the utterance, intent, and slot model means that skills can simply ask users for information verbally; researchers have identified many published skills that directly ask users for personal information~\cite{le2021skillbot, guo2020skillexplorer, edu2021skillvet}. But it is difficult to stop developers from making an intent with a slot type that captures e.g. an address and prompting users to provide this data verbally mid-conversation. Automated testing would likely struggle differentiating requests for personal data from other intents with similar definitions (e.g. naming a famous person or asking for the weather forecast at a given location), requiring manual review of every conversation tree that uses a slot type that could potentially be personal information. But even if the use of these slot types was aggressively monitored, platforms offer the ability to define custom slots that are learnt by example and can capture almost any type of information. Here the developer provides examples of what the data might look like, and the recognition model subsequently matches similar input with the slot. Custom slots have also been used in exploits: a skill gives the VA a long and silent `re-prompt', causing it to appear that an interaction has ended while the VA continues to listen for input using a custom slot crafted to match to every possible speech input~\cite{checkmarx2018alexa}. A malicious skill will then receive a transcript of everything spoken near the VA while users remain unaware that they are being recorded.

\section{Current Certification Procedures Omit Key Steps}
Beyond difficulties caused by platform architectures, the policies currently in place regarding the certification of skills are missing key elements that further allow sub-standard skills to be certified. Here we focus on three key ways that this currently manifests: (1) the lack of transparency around skill updates and lack of re-certification policies; (2) the inconsistent state of privacy policies; and (3) the lack of transparency for users about skill developers.

The first of these is perhaps the most problematic. Because skills can be hosted off-platform, there is no way for providers to know when they have been updated unless they also change their conversation model (e.g. in the Alexa Skills Kit). As a result of this, skills can silently make vast changes to their code effectively making certification voluntary and precluding a meaningful re-certification process~\cite{lentzsch2021hey, su2020you, 10.1145/3372297.3423339}; the fact that skills are able to change their operation without scrutiny in ways that could violate platform policies dramatically reduces the potential effectiveness of certification regardless of other changes to architecture and policy.

Related to this is the requirement that skills processing personal data provide a privacy policy. Developers provide URLs for their skill's policies that are checked during the certification process. Over time links to these policies can cease to function and the policies themselves can become invalid if they are not updated in line with changes to the skill. Given the issues preventing the detection of skill updates mentioned above, there is no way of knowing when a policy may have become outdated. Recent work shows that around 18\% of Alexa skills that use personal data have privacy policies that do not adequately describe data collection by the skill, with a further 25\% missing a readable policy entirely~\cite{edu2021skillvet}.

If a user, regulator, or advocacy organisation wanted to contact a developer about such an issue, how might they do so? Other than links to the privacy policy and general terms, the only information provided about a skill's developer is their name (and Google sometimes does not even provide this). This gives little to go on, commonly taking the form of a company or personal name that can be set arbitrarily by the developer~\cite{lentzsch2021hey}. Furthermore, it is not possible to effectively see other skills made by the same developer. This severely limits the ability of users to evaluate the trustworthiness of a skill before deciding whether to use it, and reduces the accountability of skill developers by making users leaving reviews or submitting `feedback' to the platform provider the only means of flagging broken or malicious skills. When users do leave reviews about dangerous or malicious skills this only affects one skill out of a developer's portfolio, and it is not clear that this process is effective; the ``Praise Me'' skill described in~\cite{edu2021skillvet} still asks for personal data verbally (bypassing the consent API) and links to a porn site instead of a privacy policy more than two years after being reported by the authors.

\section{A Provocation}
The previous sections have highlighted some of the key problems that exist within the Alexa and Google Assistant ecosystems, and potentially present in other contemporary VAs. These include difficulties stemming from platform architecture, such as the inability to audit source code, exhaustively test skills, or restrict unauthorised collection of user data, as well as issues with platform policies, including around re-certification, missing privacy policies, and lack of transparency about skill developers. The black-box nature of voice assistant platforms further means that discovering issues such as these is challenging, as researchers must construct their own automated testing infrastructure that is often in violation of Amazon and Google's terms of service, further highlighting the transparency problems with these platforms.

In response to this, we call on CUI researchers to investigate, analyse, and measure the issues highlighted in this paper. Similar efforts in smartphone apps have revealed myriad problems, many of which have subsequently been addressed. In many cases this will require collaboration between engineering, interaction, and legal/policy scholars in order to fully understand and analyse what are enormously complicated systems. 

As a starting point for future work, the next section lays out a suite of potential directions and solutions that could be taken by VAs and their associated platforms. While these are by no means exhaustive or yet fully developed, they outline a research agenda that has the potential to create more transparent platforms and reinforce user privacy.

\section{Roadmap and Future Directions}
The foundation on which a meaningful certification process must be built is the removal of loopholes that allow developers to circumvent it. The most important change in this regard would be a requirement that skills undergo re-certification when they are modified. For this to be feasible, there must also be a mechanism for detecting these changes. On smartphones, source code for apps is hosted by platform providers, making it easy to detect when an app has been updated. A similar system for voice assistant skills could involve skills being run on platform vendor's hardware---Amazon already encourages the hosting of Alexa skills on Amazon Web Services (AWS), and Google provides similar options for hosting actions on Firebase. A limitation to such an architecture is that it would not remove the potential for skills to include `dynamic content' that contravenes platform policy (i.e. when skills retrieve content from external services). This would, however, become far less likely as any violating content would have to be delivered within a skills existing conversational model. This approach, when combined with more robust user reporting mechanisms, has thus far proved viable for smartphone apps.

An implication of such a change would be that skill development would become less agile, with developers made to wait for re-certification, and a reduction in the number of non-functional or `low quality' skills, as broken updates would be rejected before deployment. The current existence of free tiers on AWS and Firebase suggests that pricing would only change for more popular skills that were forced to move to platform-provided hosting. Platforms would initially have to invest significant amounts towards increasing the capacity of their certification programs, although the long-term effects of this could be reduced via the scaling of automated methods which have access to structured conversation flows, reducing the need for human testers. Viewing these measures as introducing additional costs may prove to be further inaccurate as messaging from regulators towards online platforms increasingly takes a `self-regulate or be regulated' approach, framing increased certification measures more as a cost of doing business.

For privacy policies, it is important that these documents are reviewed regularly. Automated tools, for example, could be used to periodically verify that privacy policy links resolve to a suitable document using the same techniques as researchers are currently using to measure the issue. Given that privacy policies are comprised solely of text, an alternative option would be to host them on the platform itself. In either case, platforms should prompt developers to review their privacy policies when their skills are updated. Developer names could be constrained to real names or names of registered companies. The field would also benefit from clarification or amendment of data protection regulations to incentivise platforms to act in situations where developers shoulder the legal responsibility of being data controllers despite being heavily constrained in how they can act by the platforms that they rely on.

To facilitate the certification process itself, skill developers could provide machine-readable information on the structure of the conversation tree corresponding to each skill (e.g. the set of possible transitions from each intent). This would allow for complete coverage of skills by human and automated testers, as well as helping to prevent broken interactions when an intent fires unexpectedly by enforcing conversation structures at runtime. Similar methods have led to the development of more complex verification mechanisms on other platforms, such as Android app manifests which define a contract between the functionality of the app and the permissions requested. These are used to automatically enforce the correct use of privileged resources when apps are running. However, systematising conversation trees for skills would further entrench the current utterance, intent, slot paradigm used by many voice assistants, making it more difficult for platforms to introduce other ways of interacting with skills. As such, new CUI paradigms might appear as separate products rather than being housed under the same umbrella as current skills.

Finally, platforms could constrain the data types made available through slots attached to conversational intents, which have previously formed the basis for attacks~\cite{checkmarx2018alexa}. These options could be restricted to a subset of approved developers or accept a limited number of sample utterances to prevent every potential input matching the slot. This could have a big impact for CUI design, limiting the extent to which developers can create specialist applications and use existing platforms in new and innovative ways. The result could be a homogenisation of skills on mainstream platforms, pushing others to smaller services. A potential problem here is the hegemony of major tech companies when it comes to the advanced machine learning models that power speech recognition, natural language processing, and speech synthesis, especially as more organic modes of interaction are likely to require more sophisticated processing.

\section{Conclusion}
It is inevitable as voice assistant platforms evolve and develop that there will be growing pains, especially as the societal and regulatory landscape around internet enabled devices continues to change. In this provocation paper we have collected together the key problems facing major platforms and called on members of the research community to investigate and document them. Exploration of the solution space will also be required in order to develop platforms that safeguard their users. Finally, the paper also sets out the potential implications for the design of voice assistants and CUIs more broadly should platforms act in the ways suggested in an attempt to mitigate these problems. Voice assistants, like other online platforms, are shaped by the policies that they choose to enforce, and third-party designers and developers will look closely at action taken by household voice assistant vendors, who are responsible for the most widely used CUIs in the world.

\begin{acks}
This work is funded by the Secure AI Assistants project via Grant EP/T026723/1 from the UK Engineering and Physical Sciences Research Council (EPSRC).
\end{acks}

\balance

\bibliographystyle{ACM-Reference-Format}
\bibliography{refs}

\end{document}